\documentclass[12pt]{iopart}

\usepackage{iopams} 
\usepackage{amssymb}
\usepackage{hyperref}
\hypersetup{colorlinks=true,linktocpage=true}
\usepackage{framed,color}
\definecolor{shadecolor}{rgb}{1,0.8,0.3}

\usepackage{color} 
\usepackage{graphicx}
\usepackage[T1]{fontenc} 

\begin{document}

\title{Population-expression models of immune response}

\author{Sean P Stromberg$^{1,3}$, Rustom Antia$^1$, and Ilya Nemenman$^{1,2}$}

\address{$^1$ Department of Biology,} \address{$^2$ Department of
  Physics, and Computational and Life Sciences Initiative\\ Emory
  University, Atlanta, Georgia 30322, USA}
\address{$^3$ Department of Physics, University of California, Santa Barbara, CA 93106, USA}

\ead{stromberg@physics.ucsb.edu, \{rantia, ilya.nemenman\}@emory.edu}

\begin{abstract}  
  The immune response to a pathogen has two basic features. The first
  is the expansion of a few pathogen-specific cells to form a
  population large enough to control the pathogen. The second is the
  process of differentiation of cells from an initial naive phenotype to an effector
  phenotype which controls the pathogen, and subsequently to a memory phenotype that is maintained and 
  responsible for long-term protection. The expansion and the
  differentiation have been
  considered largely independently. Changes in cell populations are
  typically described using ecologically based ordinary differential
  equation models. In contrast, differentiation of single cells is
  studied within systems biology and is frequently modeled by
  considering changes in gene and protein expression in individual
  cells. Recent advances in experimental systems biology make
  available for the first time data to allow the coupling of
  population and high dimensional expression data of immune cells
  during infections. Here we describe and develop {\em
    population-expression} models which integrate these two processes
  into systems biology on the multicellular level. When translated
  into mathematical equations, these models result in
  non-conservative, non-local advection-diffusion equations. We
  describe situations where the population-expression approach can
  make correct inference from data while previous modeling approaches
  based on common simplifying assumptions would fail. We also explore
  how model reduction techniques can be used to build
  population-expression models, minimizing the complexity of the model
  while keeping the essential features of the system. While we
  consider problems in immunology in this paper, we expect
  population-expression models to be more broadly applicable.
\end{abstract}

\pacs{05.45.-a, 87.10.Ed, 87.18.Vf, 87.19.lx, 87.19.xw}
\vspace{2pc}
\noindent{\it Keywords}: Immunology, Heterogeneity, Model Reduction, Population-Expression, Systems Biology

\submitto{\PB}
\maketitle

\section{Introduction}
\label{sec:Intro}
\begin{figure}[htbp]
	\centering
	\includegraphics[scale=1.0]{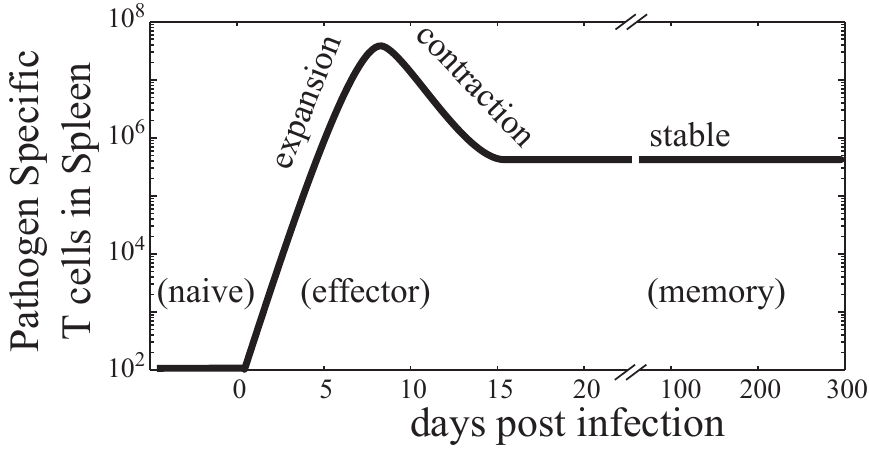}
        \caption{Schematic illustration of a typical CD8 T cell
          response to an infection.  The plot shows the enormous
          changes in numbers of pathogen-specific CD8 T cells during
          the course of infection, as well as changes in cell
          phenotype.  The response has three phases which correspond
          to population expansion, contraction and stability.
          Differentiation results in changes in the phenotype of cells
          from naive to effector and memory.  Typically this type of a
          response is described by ordinary differential equations
          that govern changes in populations of cells having naive,
          effector and memory phenotypes. \label{fig:Expansion}}
\end{figure}

The central feature of the adaptive immune system is the ability to
respond to a broad range of pathogens, including emerging threats
never before encountered, without mounting responses to the native
tissues of the body \cite{Perelson:1997vk}. This dynamic is explained
by the {\em clonal selection theory}, which underlies our understanding of
immunology. This theory postulates that we begin with a very diverse
population of immune cells (lymphocytes), with each lymphocyte having
a unique and fixed specificity. Consequently the number of lymphocytes
specific for a given pathogen is very small.  Following infection these pathogen-specific lyphocytes undergo rapid division (clonal expansion)
and differentiation into effector cells, which are able to
control the pathogen. Following clearance of the pathogen some of
these lymphocytes differentiate into memory cells, which are maintained
for extended periods and account for long-term protection. The
clonal selection theory describes the generation of the T cell and B
cell responses. In Fig.~\ref{fig:Expansion} we show a schematic of
clonal selection for a typical CD8 T cell response to a viral
infection.

The enormous changes in population sizes suggested that, as in
ecology, ordinary differential equation (ODE) models of the
populations would prove useful to understanding the immune response
\cite{nowak_virus_2001,perelson_modelling_2002,antia_role_2005}. In
these models cells are restricted to a few distinct phenotypes with
division, death, and transition rates between the phenotypes to
describe the dynamics. The models typically ignore how the systems
biology on the cellular scale governs the rate laws in the models on
the population scale. While such models have proven useful in
addressing a number of population level questions, they have their
limitations. For the approach to work well, phenotypic states must be
well resolved and the transitions between them must be rapid.

\begin{figure}[htbp]
	\centering
	\includegraphics[scale=1.0]{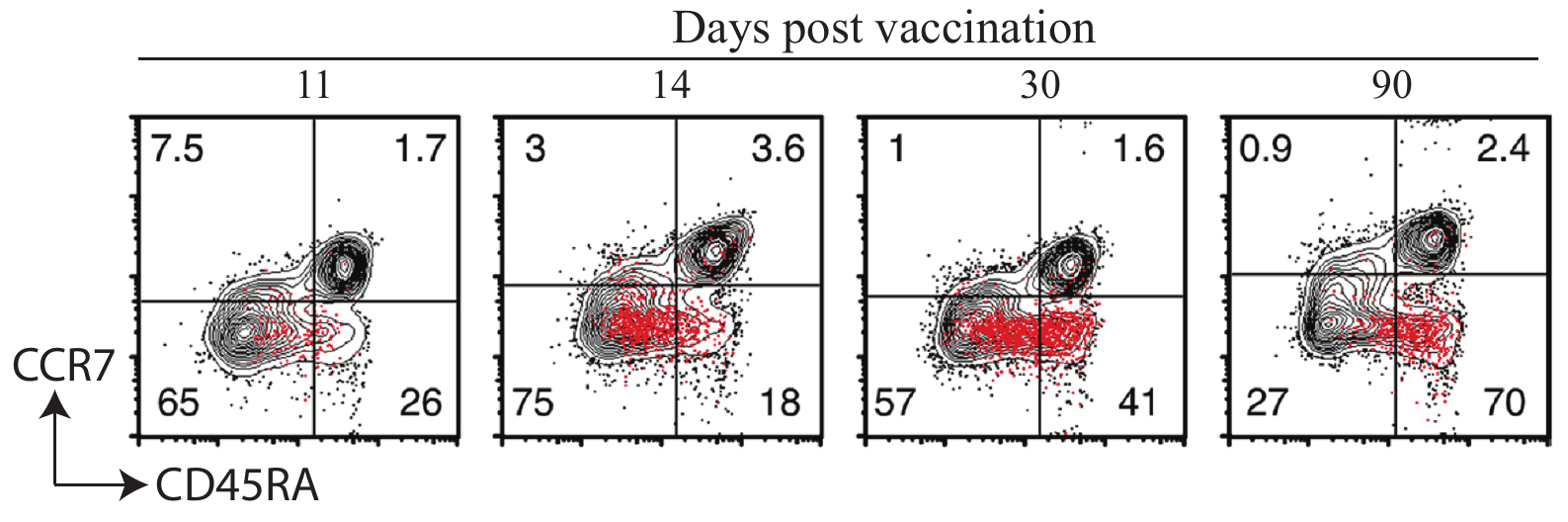}
        \caption{The differentiation of human CD8 T cells following
          yellow fever vaccination. These flow cytometry plots show
          the population of antigen specific CD8 T cells (red)
          responding to the vaccination differentiating from CD45RA
          negative to positive while expanding and then contracting in
          number. This transition is associated with the transition
          from effector memory to central memory. Reproduced with
          permission from
          \cite{akondy_yellow_2009} {\it (Copyright 2009. The American Association of Immunologists, Inc.)} \label{fig:cytometry}}
\end{figure}

Fig.~\ref{fig:cytometry} presents data capturing the dynamics of T cells 
obtained via flow cytometry. This figure shows the density of CD8 T
cells following a yellow fever vaccination plotted as a function of
two surface expressed molecules (CD45RA, a signaling molecule that
regulates antigen receptor signaling, and CCR7, a molecule which aids
in trafficking of T cells to lymph-nodes)
\cite{akondy_yellow_2009}. The population gradually transitions from
CD45RA low to high during the contraction and memory phases. This
figure illustrates one problem with ODE models of multicellular
population dynamics: How does one unambiguously partition data into
distinct phenotypes when there is considerable heterogeneity or
gradual transitions? This ambiguity gives rise to subjectivity and
quantitative disagreement between labs in the analysis of
immunological data \cite{maecker_standardizing_2012}.

The flow of populations as they differentiate
(Fig.~\ref{fig:cytometry}) is governed largely by the systems biology
of the cells
\cite{chakraborty_pairing_2010,goldstein_mathematical_2004,goldstein_kinetic_2008}. (While
the term {\em systems biology} has been used very broadly, in this
manuscript we adopt the most common usage, referring to models of
chemical reaction networks typically within single cells or
homogeneous cell cultures \cite{Ferrell:2009}.) Typical systems
biology models consist of ODEs or stochastic differential equations
that model reaction rates between chemical species, providing a finer
resolution of phenotypic states.

While population models loose accuracy in not considering the chemical
scale, systems biology models have contrasting limitations resulting
from omission of the population dynamics. Typically, the analysis and
parameter estimation of differentiating populations has been performed
on time scales where division is negligible \cite{munsky2009}. On
longer time scales, population dynamics and systems biology are
coupled and must be considered together. The expression levels of gene
products control cell division and death rates. In their turn, cell
division and death rates change the number of cells in various
phenotypic states and hence shape the expression profiles of
populations. Additionally the process of cell division dilutes
expression and can generate spurious correlations between expressed
chemicals. Clearly, modeling immune system dynamics requires an
integrated approach, combining population dynamics and systems
biology.

One way to do this is to conceptualize each flow cytometry data set as
samples from a density in a multidimensional cellular configurational
space, where each dimension denotes the quantity of a specific
chemical. Individual cells would trace out trajectories in this
configurational space as they differentiate. Unfortunately {\em in
  vivo} single cell longitudinal data is difficult to obtain, and for
dividing cells the term longitudinal is undefined.  Thus instead of
tracking cells over time, one can focus on tracking populations, or
distributions of cells in the configurational space. This can be done
using partial differential equations (PDEs) and related mathematical
concepts, an approach gaining popularity in theoretical immunology
\cite{yates_combining_2004,Banks2010,stromberg_role_2012}. We refer 
to the dynamics of chemical expression (gene, protein, metabolite, etc.)
in a dynamic population as the {\em population-expression}, and models of the 
population-expression as {\em population-expression models}.

Such population-expression models circumvent our inability to
define clean cellular phenotypic states. They remove the inherent
subjectivity in phenotype discrimination
\cite{maecker_standardizing_2012}, and they remove the need to
incorporate additional phenotypes to better fit models to data. They
integrate the within cell stochastic chemical kinetics into models of
the population dynamics. Ultimately, they allow analysis of the
diversity of protein expression within populations, how it changes
with time, and how the diversity is affected by selection.

The main goal of this paper is to introduce such population-expression
modeling, explain utility of the approach in the context of toy
models, and discuss the methodological developments needed for
practical applications of the ideas. To achieve this, we first
introduce a formalism for population-expression
models using PDEs, and non-local PDEs. Following this we provide a number of examples of
population-expression models, illustrating where ecological based ODE
models succeed and fail, how cell division dilutes chemical
quantities, where single-cell analyses fail to describe the
population, and how we can infer from data which chemicals may be
drivers in regulatory networks. We end with a critical look at some of
the key problems arising when we confront population-expression models
with ever increasing dimensionality of experimental datasets.

\section{Population-expression approach: PDE formulation}
Instead of predefining a limited number of cell phenotypes, our
population-expression approach takes the abundance of cells with
different chemical states as the dynamical variable. We denote by
$\rho(\vec{A},t)$ the density of cells at time $t$, with internal
biochemical expressions (internal states) of $\vec{A}$.

To describe how $\rho(\vec{A},t)$ changes with time, we first consider
how a single cell moves in the configurational space of $\vec A$
values. Denoting the set of differential equations that describe the
changing chemical quantities within a single cell by 
\begin{equation}
  \frac{d \vec A}{dt}=\vec \gamma(\vec
  A),\label{chemkin} 
\end{equation}
the abundance $\rho$ flows according to the vector field denoted by
$\vec \gamma (\vec A)$. A number of techniques exist to translate from the single cell
model to a population model \cite{metz_dynamics_1986}. In the accompanying Supplementary Materials we
provide two such contrasting derivations, one more common to the fluid-dynamics
community (based on the divergence theorem), and the other more common to statistical
physics and systems biology (based on the chemical master equation). These techniques 
have identical results, generating an advection equation describing how the density 
changes according to the vector field $\vec \gamma(\vec A)$:
\begin{equation}
  \frac{\partial \rho(\vec A,t)}{\partial t} = -{\vec{\nabla}} \cdot\left[
    \vec \gamma(\vec A)\rho(\vec A, t)\right].\label{eq:advec}
\end{equation}
The quantity in the square brackets denotes the total flux of cells
changing in expression level as they move through the configurational space, 
and $\vec{\nabla}$ defines the divergence operator, a vector of partial derivative operators ($\partial/\partial A_1,\partial/\partial A_2,\ldots$). The formulation is valid for arbitrary dimensionality, and the examples in the following sections use either one or two dimensions for simplicity. 

Incorporating population dynamics into these equations can be done
with additional terms for cell death and sources of new cells:
\begin{equation}
\frac{\partial \rho(\vec A,t)}{\partial t} = -\vec \nabla \cdot\left[ \vec
  \gamma(\vec A)\rho(\vec A, t)\right]-\nu(\vec A)\rho(\vec
A,t)+\Gamma(\vec A).
\label{eq:pop}
\end{equation}
Here $\nu(\vec A)$ denotes a cellular death rate that is a function of
the chemical concentration and $\Gamma(\vec A)$ is an influx of new
cells entering the system in a chemical state $\vec{A}$.

Cell division can be included by adding nonlocal terms to
Eq.~(\ref{eq:pop}). For example, if in a symmetric cell division, all
chemicals in the cell are split equally between the two daughters, we
have
\begin{eqnarray}
   \frac{\partial \rho(\vec A,t)}{\partial t} &=& -\vec \nabla \cdot\left[ \vec
    \gamma(\vec A)\rho(\vec A, t)\right]-\mu(\vec A)\rho(\vec A,t)
+2^d
  \cdot 2 \mu(2\vec A)\rho(2\vec A,t)\nonumber\\&&-\nu(\vec A)\rho(\vec
A,t)+\Gamma(\vec A).\label{eq:dilution}
\end{eqnarray}
Here $\mu(\vec{A})$ is the division rate, which we assume depends on
the cell age and other properties only implicitly through the
instantaneous state of the cell, $\vec{A}$. In this equation, cells with 
chemical quantity $\vec A$ are removed from the abundance at $\vec A$ as they divide
with rate $\mu(\vec A)$. Separately, each cell dividing at abundance  $2\vec A$ is adding 
two cells to the abundance at $\vec A$. The factor of $2^d$ arises from
a subtlety of the non-local calculus. Division adds {\it to} an infinitesimal volume of the 
space, bounded in each dimension by $(A_i,A_i+\delta A_i)$ where $\delta$ is an infinitesimal quantity. The cells however are coming {\it from} a region with boundaries $(2 A_i, 2
A_i+2\delta A_i)$, which is twice the width in each of the $d$ dimensions. 
This equation can also be modified to describe dilution in
asymmetric cell division.

It is frequently the case that non-locality gives rise to integro-differential equations. 
If we incorporated partitioning noise into our equation, it would generate 
an integral term as new cells would enter the population at $\vec A$ from a range
of values centered around $2\vec A$. When considering systems with small numbers of molecules, this
approach is an important extension. With large numbers of molecules, the 
relative variation is small, and partitioning noise can be neglected.

Finally we can very naturally incorporate the stochastic fluctuations
resulting from the chemical dynamics \cite{Elowitz:2002,Raser:2004}.
This is typically done by constructing a chemical master equation, and
expanding in small relative fluctuations
\cite{kampen_stochastic_2007}. An example derivation is provided in the 
Supplementary Materials. Expanding the chemical master equation to lowest order gives
Eq.~(\ref{eq:advec}). The expansion of the chemical master equation to 
next highest order results in a nonlocal analogue of the Fokker-Planck 
equation, which spreads the
population in the $\vec A$ space due to stochasticity of the intrinsic
chemical processes: 
\begin{eqnarray}
 \frac{\partial \rho(\vec A,t)}{\partial t} &=& -\vec \nabla \cdot\left[ \vec
    \gamma(\vec A)\rho(\vec A, t)\right]-\mu(\vec A)\rho(\vec A,t)+2^d
  \cdot 2 \mu(2\vec A)\rho(2\vec A,t)\nonumber\\&&-\nu(\vec A)\rho(\vec
A,t)+\Gamma(\vec A)+\vec \nabla \cdot (\mathbf{D}(\vec A)\nabla\rho(\vec A,t)).\label{eq:FP}
\end{eqnarray}
Here $\mathbf{D}(\vec A)$ is a diffusion tensor. The advection dynamics 
becomes advection-diffusion dynamics with the incorporation of 
within-cell stochasticity.
Such approaches to modeling fluctuations in single cells are now
commonplace in molecular systems biology \cite{Paulsson:2004}, and
many efficient simulation and analysis algorithms have been developed
\cite{munsky_modeling_2012}. 

As in systems biology, in population-expression models some
state variables may remain discrete. For example the state of
transcription factor binding may be best described by a binary {\em
  on/off} variable, or compartmental spacial dependence could be
incorporated into the model (e.g. lung, spleen, etc.). In these cases
we typically describe multiple coupled densities $\rho_i(\vec A,t)$,
with population-expression dynamics, Eq.~(\ref{eq:FP}), for each
density and with terms that couple the equations through transitions
between the states, such as $\sum_i k_j \rho_j(\vec A, t)$.
Of possibly high relevance to the current work, cross-sectional flow
cytometry samples from cellular populations at different time points
have been used to infer parameters of chemical reaction rates $\vec
\gamma$ \cite{munsky2009}. The population-expression approach differs
from these analyses by incorporating the effects of proliferation,
cell death, and dilution by cell division. We show below that these
effects can substantially bias the resulting expression profile of a
population. 

Note that, for much of this paper, we assume that $\rho(\vec A,t)$ can
be measured: that the number of samples is large enough so that
inference of $\rho$ is not a hard task. This breaks down if $d=\dim
\vec{A}\gg1 $. We will discuss this case in
Sec.~\ref{model-reduction}. Similarly, we assume that
population-expression equations are sufficiently low-dimensional to be
numerically solvable. When this is not the case, Monte-Carlo
simulations might be needed, and we briefly touch on this topic in the
{\em Discussion}.

\section{Population-expression approach: Examples}
In this section, we use the population-expression approach to model
simple processes of relevance to different aspects of immune
dynamics. The examples illustrate the inadequacy of single cell systems
biology (expression) and ecological based ODE (population)
modeling approaches.

\subsection{Ecological based ODE model failure: slow expression dynamics}
\label{sec:Ex1}
Ecological based ordinary differential equation models of phenotypical population
dynamics work well only when phenotypes are sharply defined and
transitions between them are rapid. This is not always the case.
Consider, for example, a transition between phenotypes that occurs
when an internal state has changed, but the observables take time to
reach their characteristic values for this new state. For example, a
good measure of the phenotypic state of a cell may be the binding of
transcription factors (TFs) to DNA, which is possible but not easy
to measure \cite{berger_universal_2009}. On the other hand, we
routinely measure expression levels of protein using flow
cytometry. These levels are typically controlled by transcription
factor binding, but changes in protein expression lag behind changes
in TF binding. Thus the dynamics of switching observed in flow
cytometry data may be non-trivial.

Here we model cells having a discrete state denoting transcription
factor binding (``off'' or ``on''), and a continuous variable $A$ for
expression level. Off cells can switch to the on state with the rate
$k$, and the dynamics of $A$ is given by $dA/dt=\gamma_{\rm off/on}$,
where $\gamma_{\rm off/on}$ depends on the state.  Namely, the
chemical $A$ has two possible production rates: $\alpha_{\rm off}$,
and $\alpha_{\rm on}$. In both states there is the same degradation
rate $\beta$. This kinetics may correspond, for example, to the
expression and decay of mRNA or protein if mRNA levels equilibrate
quickly in comparison to the protein dynamics. We consider the cells
in the two states separately: $\rho_{\rm off}(A,t)$ is the density of
cells in the off state with expression level $A$, and $\rho_{\rm
  on}(A,t)$ are the cells in the on state. The population-expression
equations are:
\begin{eqnarray}
	\frac{\partial \rho_{\rm off}(A,t)}{\partial t} &=& -
        \frac{\partial}{\partial A} [(\alpha_{\rm off} - \beta
        A)\rho_{\rm off}(A,t)]+\nonumber\\
	&&\frac{1}{2} \frac{\partial^2}{\partial
          A^2}[(\alpha_{\rm off} + \beta A)\rho_{\rm off}(A,t)] -
        k\rho_{\rm off}(A,t)\label{eq:state1},\\
	\frac{\partial \rho_{\rm on}(A,t)}{\partial t} &=& -
        \frac{\partial}{\partial A} [(\alpha_{\rm on} - \beta
        A)\rho_{\rm on}(A,t)]+\nonumber\\
	&&\frac{1}{2} \frac{\partial^2}{\partial
          A^2}[(\alpha_{\rm on} + \beta A)\rho_{\rm on}(A,t)] +
        k\rho_{\rm off}(A,t)\label{eq:state2}.
\end{eqnarray}
Similar models for single cells in equilibrium
\cite{iyer-biswas_stochasticity_2009,mugler_spectral_2009}, and even
off-equilibrium for simpler cases \cite{Ramos:2011wn}, have been
solved exactly. Here we analyze this system numerically in the
non-equilibrium context. We solve these equations with a method of
lines integration with a finite differencing approximation for $A$
derivatives, and Matlab ODE45 routine for integrating forward in time.

\begin{figure}[b]
	\centering
		\includegraphics[scale=1.0]{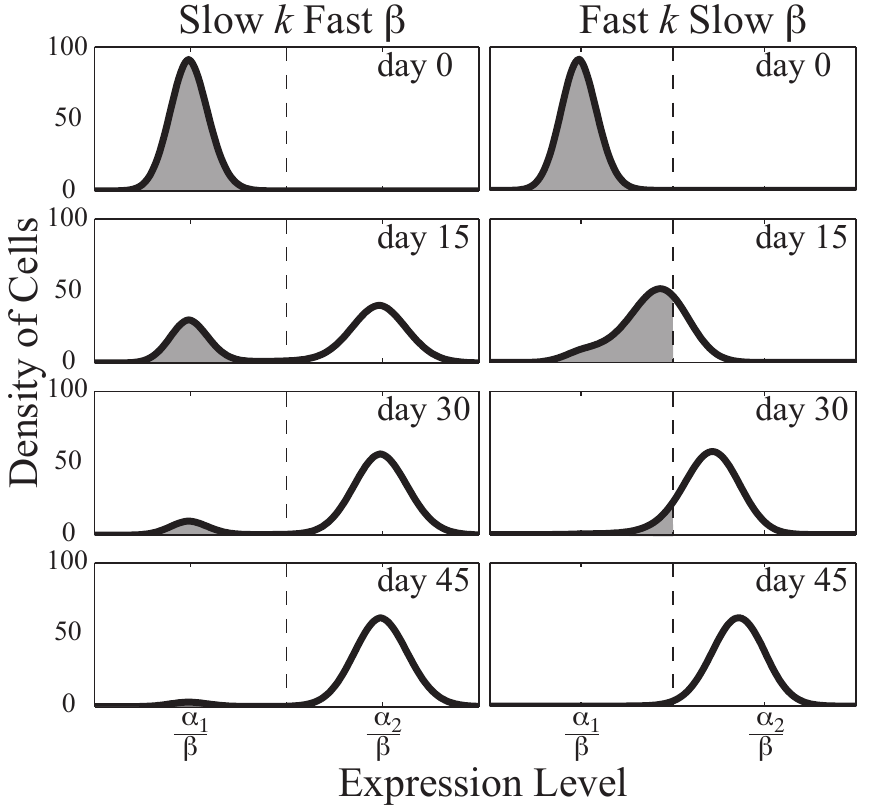}
                \caption{Expression profiles at multiple time points
                  for slow state switching / fast protein production
                  (left column) and fast switching / slow protein
                  production (right column). The dashed line
                  represents the half-way point between the steady
                  state protein concentrations in the two states. ODE
                  models typically count the number of cells above and
                  below a threshold like this one to consider
                  transition rates. The parameters of the two models
                  are such that the steady state values and the
                  overall protein relaxation times are the same. The
                  shaded region corresponds to the number of
                  low-expressing cells plotted in
                  Fig.~\ref{fig:ODECorrespondence}. In the simulation
                  on the left: $\alpha_1$ = 94.5 copies/day,
                  $\alpha_2$ = 190 copies/day, $\beta$ = 1.0
                  day$^{-1}$, $k$ = 0.075 day$^{-1}$; while on the
                  right: $\alpha_1$ = 5.0 copies/day, $\alpha_2$ =
                  10.0 copies/day, $\beta$ = 0.05 day$^{-1}$, $k$ =
                  0.2 day$^{-1}$.}
	\label{fig:4}
\end{figure}
Fig.~\ref{fig:4} plots numerical solutions of $\rho(A,t)=\rho_{\rm
  off}(A,t)+\rho_{\rm on}(A,t)$, defined by
Eqs.~(\ref{eq:state1},~\ref{eq:state2}), for two contrasting pictures
of differentiation. The left panels shows infrequent TF switching with
rapid protein expression ($k\ll\beta\langle A\rangle$). In this case
the protein concentration in each cell tracks its transcriptional
state well, phenotypes are well defined, and an ODE model describing
switching between them works well. The right panels in
Fig.~\ref{fig:4} represent the case when TF switching is rapid, but
change in protein expression is gradual. The initial and final states
are identical to the scenario on the left. The gradual protein
expression gives a large density of cells with intermediate protein
expression on day 15, and no well-resolved phenotypes.

\begin{figure}[htbp]
	\centering
		\includegraphics[scale=1.0]{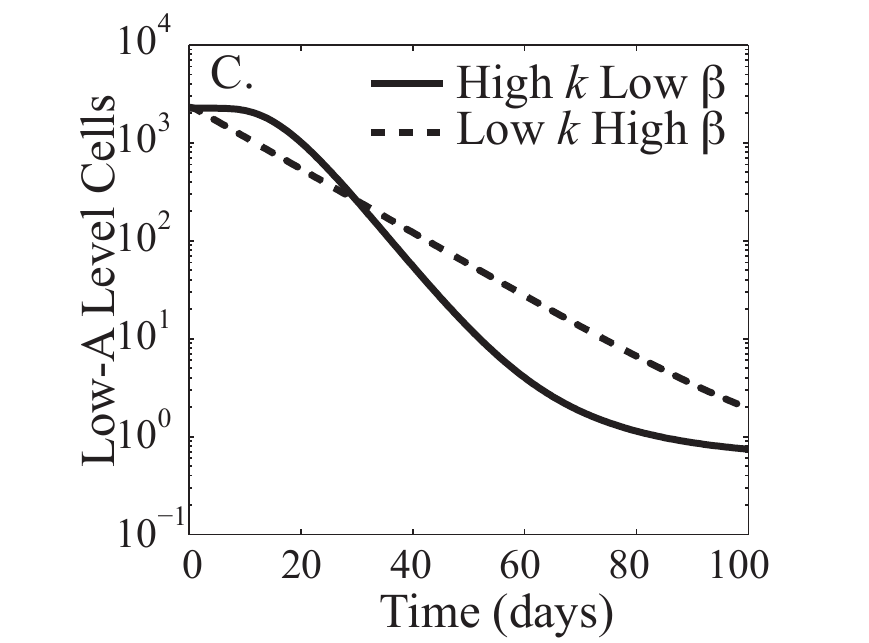}
                \caption{The total number of cells in the
                  low-expression phenotype (as defined by the cutoff
                  in Fig.~\ref{fig:4}) for the two scenarios. When the
                  expression level equilibrates rapidly after the
                  transcription factor is bound (dashed line), the
                  system has a single characteristic decay time scale,
                  and the number of cells in the low state can be
                  modeled with a single ODE. When the protein
                  expression dynamics are slow to respond (solid
                  line), the decay of the population in the low state
                  is non-exponential.}
	\label{fig:ODECorrespondence}
\end{figure}

Dashed lines in Fig.~\ref{fig:4} define low and high expressing
phenotypes, as is typical in the analysis of flow cytometry data. The number
of cells in the low expressing phenotype is shown as a function of
time in Fig.~\ref{fig:ODECorrespondence} for both scenarios. For
rare switching, modeling the system with two phenotypes with
population sizes $X_1$ and $X_2$, respectively, as
\begin{eqnarray}
\frac{dX_1}{dt}=  -k X_1,\nonumber\\
\frac{dX_2}{dt}=  +k X_1,
\end{eqnarray}
produces great fits to the data. In contrast, two-state modeling for
the slow protein expression case is inaccurate.

To model the data with
ODEs and discrete states, several approaches could be taken. 
As is common in immunology \cite{saini_regulation_2010} one could
introduce sub-phenotypes, partitioning the cells into $n>2$ domains by
some  predefined thresholding of their expressions, such that
\begin{eqnarray}
\frac{dX_1}{dt}=  -k_1 X_1,\nonumber\\
\frac{dX_i}{dt}=  -k_i X_i+k_{i-1}X_{i-1},\quad i=2,\dots,n-1,\nonumber\\
\frac{dX_n}{dt}=  +k_{n-1} X_{n-1}.\label{eq:odephenotypes}
\end{eqnarray}
One would then optimize the parameters $k_i$ to produce the best fit
to the data. This approach also has its limitations. The steady state
distribution given by Eqs.~(\ref{eq:state1}) and (\ref{eq:state2}) has
a width, while Eq.~(\ref{eq:odephenotypes}) has a steady state where
all cells are within the $X_n$ partition. Any overlap between the
steady state distribution and the $X_{n-1}$ state will not be resolved
by such a model. Additionally this method introduces spurious
phenotypes having little to do with the underlying biology.
Alternatively, one could make the transition rate $k$ a function of
time $k(t)$. Like the previous case, this technique describes the
data, but provides little insight into the biology of the system.

\subsection{Failure of single cell systems biology: Cell division}
The models presented here are constructed based on chemical number
rather than concentration. This gives correspondence with fluorescence
experiments and enables accurate estimation of stochastic
effects. Upon cell division we must divide the contents of a cell in
half (assuming symmetric cell division). This gave us the non-local
PDE in Eq.~(\ref{eq:dilution}). Such nonlocal partial differential
equations are uncommon and most computational tools are ill-equipped
to deal with them.  The use of finite difference, finite element, and
spectral methods in solving these types of equations has been studied
in a series of papers
\cite{mantzaris_numerical_2001a,mantzaris_numerical_2001b,mantzaris_numerical_2001c}. For
large dimensional systems, Monte-Carlo integration can provide a more
efficient numerical solution. In these examples we use finite
difference methods.

\subsubsection*{{\bf Dilution of a dye:}}
As a simple example of dilution by division, consider a dye such as
CFSE or BrdU. These dyes are used to measure cell division rates in
vivo and are frequently used in studying the cellular dynamics of
immune responses.  These dyes are not produced by the cells and are
degraded slowly, yielding $\gamma = 0$. This removes the advection
term in Eq.~(\ref{eq:dilution}) yielding:
\begin{equation}
	\frac{\partial \rho(A, t)}{\partial t} =  - \mu \rho(A, t) + 4 \mu \rho(2A, t).\label{eq:dye}
\end{equation}

\begin{figure}[htbp]
	\centering
		\includegraphics[scale=1]{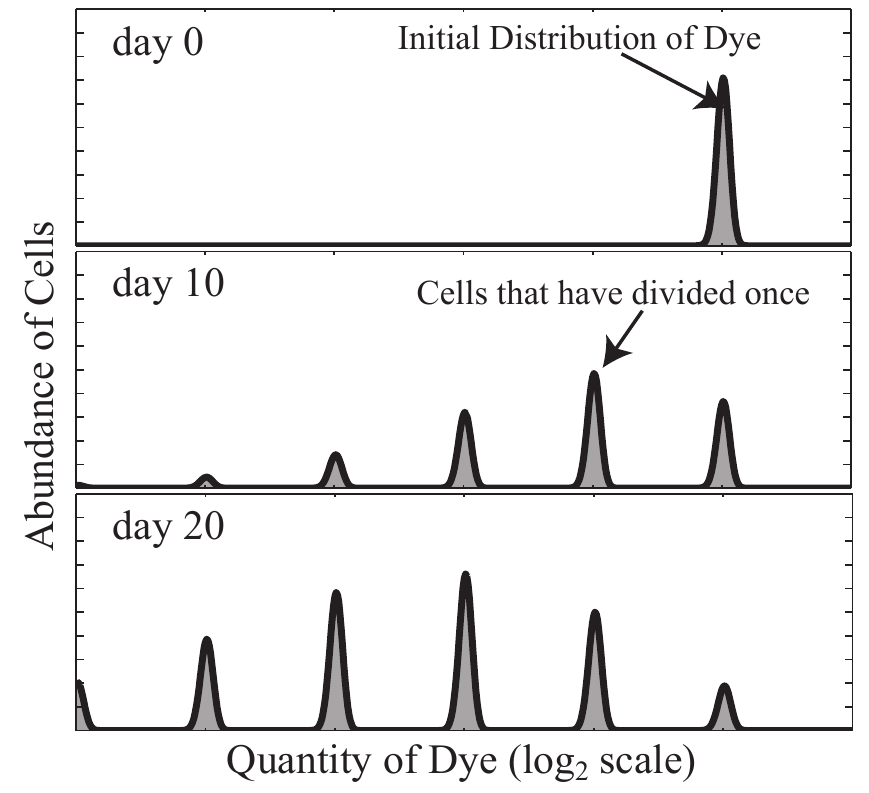}
                \caption{Dilution of dye upon cell division. The
                  rightmost peak is the initial undivided cells. Each
                  peak to the left is cells that have gone through an
                  additional division. Here $\mu$ = 0.07 day$^{-1}$.}
	\label{fig:dye}
\end{figure}

For a dye that initially has a narrow Gaussian distribution in cells, we
have the output shown in Fig.~\ref{fig:dye}. This system has been well
described using ODE models
\cite{revy_functional_2001,de_boer_estimating_2006}, with a single ODE
for the number of cells in each peak. We note that, for brevity, we
are using a model with exponentially distributed division times. For
rapidly dividing cells, more detailed models of cell cycle provide
greater accuracy \cite{de_boer_estimating_2006,Banks2010}.

\subsubsection*{{\bf Dilution and homeostasis:}}

\begin{figure}[htbp]
	\centering
		\includegraphics[scale=1]{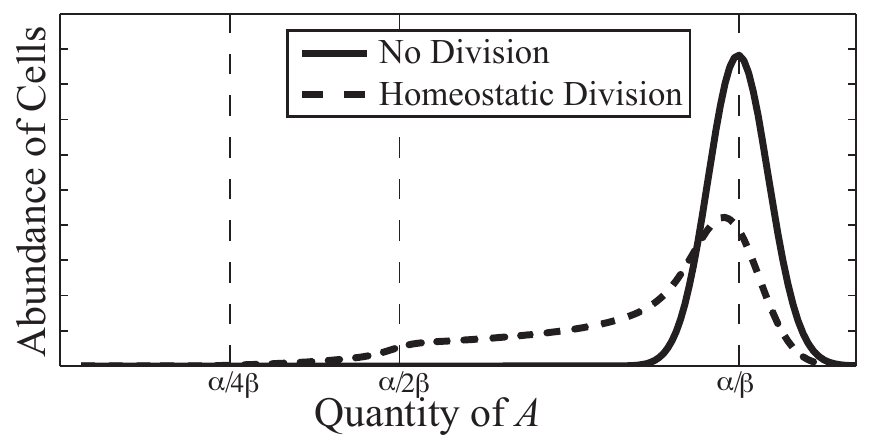}
                \caption{Stable distributions from a non-dividing
                  population and one that is undergoing homeostatic
                  division. Homeostatic division biases the
                  population-expression, reducing the mean and
                  increasing the width. The effects of cell division
                  on the population-expression are greater the more
                  rapid the division rate is. Parameter values were
                  $\alpha$ = 200 copies/day, $\beta$=0.4 day$^{-1}$,
                  $\mu$ = 0.07 day$^{-1}$.}
	\label{fig:divisionbias}
\end{figure}

For a chemical that is produced in the cell, division can bias the
population-expression. Fig.~\ref{fig:divisionbias} shows an example of
this effect. Here we have simulated two populations of cells, one
not-dividing (solid) and one undergoing homeostatic division (cell
death and division rates are equal, dashed curve). These curves are
stationary distributions generated by the equation:
\begin{eqnarray}
	\frac{\partial \rho(A, t)}{\partial t} &=&
        -\frac{\partial}{\partial A} [(\alpha - \beta A)\rho(A,t)] +
        \nonumber\\
&&\frac{1}{2} \frac{\partial^2}{\partial
          A^2}[(\alpha + \beta A)\rho(A,t)]- 2\mu \rho(A, t) + 4 \mu \rho(2A, t), \label{eq:dilution4}
\end{eqnarray}
where we have also included the stochastic effects of the chemical
dynamics.  To keep the system from growing, we have cell death rate
equal to division rate, giving an extra factor of 2 in the second to
last term.

Fig.~\ref{fig:divisionbias} shows stable distributions for this system
with and without $\mu=0$. As we can see, cell division biases the
distribution, reducing the mean and increasing the width. In general,
the more rapid the division, the more exaggerated the effects. If the
division rate exceeds the chemical degradation rate $\beta$, the
stable distribution is very different from what is seen here, and is
centered close to $A=0$.

Statistical deviations resulting from cell division have been studied
in previous work
\cite{swain_intrinsic_2002,rosenfeld_gene_2005,rosenfeld_fluctuation_2006}. This
type of noise is typically considered {\em extrinsic noise}
\cite{swain_intrinsic_2002}. It is also often approximated as a local
and continuous process and incorporated into chemical decay terms
\cite{ozbudak_multistability_2004}, a modeling choice which omits many
of the effects illustrated in this section.

\subsubsection*{{\bf Dilution and expansion:}}

\begin{figure}[htbp]
	\centering
		\includegraphics[scale=1]{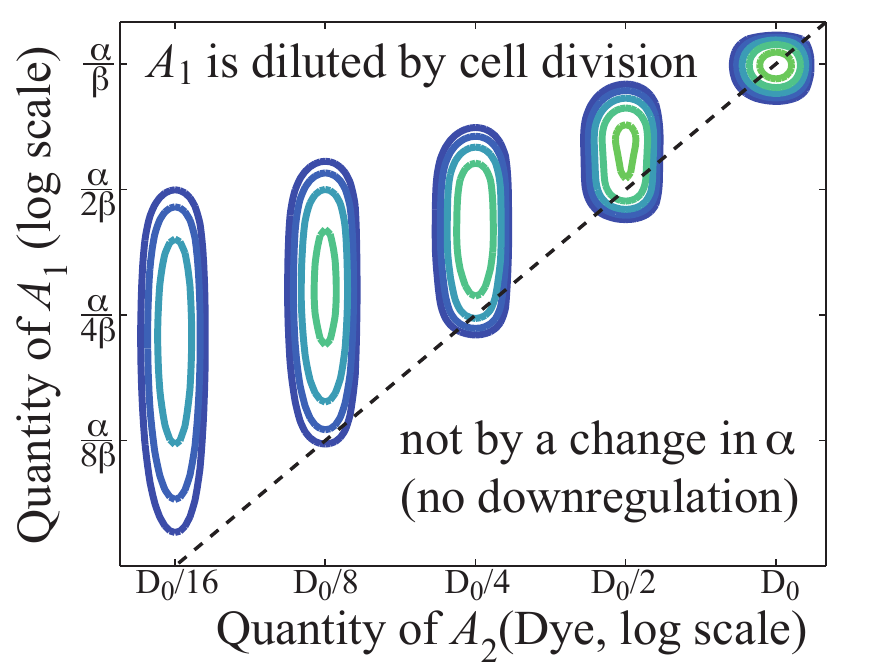}
                \caption{Simulation of a population of cells on day 7
                  for a two dimensional system with chemical $A_1$
                  obeying chemical rate law $dA_1/dt=\alpha-\beta
                  A_1$, and $A_2$ representing a dye. Contours are
                  spaced logarithmically and we have included
                  stochastic effects of $A$ expression. The population
                  initially had $A_1$ distributed at equilibrium for a
                  non-dividing population. At day 0 the population
                  began dividing which dilutes $A_1$ and $A_2$, though
                  $A_1$ is produced in the cell giving a vertical
                  spread. This simulation has correspondence with a
                  resting population of lymphocytes that is dyed and
                  then stimulated by an infection on day 0 resulting
                  in rapid expansion. Thought there is no change in
                  the production rate $\alpha$ dilution gives a
                  reduction in expression. Population-expression
                  models can discriminate between reduction in
                  expression resulting from a change in chemical
                  dynamics and this simple dilution. Here $\mu$ = 0.09
                  day$^{-1}$, $\alpha$ = 43 copies/day, $\beta$ = 0.08
                  day$^{-1}$.}
	\label{fig:dilutiondownreg}
\end{figure}

Fig.~\ref{fig:dilutiondownreg} shows a simulation of a bivariate
process where the vertical axis represents a chemical $A_1$ that is
produced by the cell, as in Eq.~(\ref{eq:dilution})
(Fig.~\ref{fig:divisionbias}), and the horizontal axis represents a
dye concentration $A_2$, with dynamics as in Eq.~(\ref{eq:dye})
(Fig.~\ref{fig:dye}). Here the population is expanding rather than
undergoing homeostatic division. The equation describing the dynamics
of the system is:
\begin{eqnarray}
\fl	\frac{\partial \rho(A_1,A_2, t)}{\partial t} &=&  -\frac{\partial}{\partial A_1} [(\alpha - \beta A_1)\rho(A_1,A_2,t)] + \frac{1}{2} \frac{\partial^2}{\partial
          A^2_1}[(\alpha + \beta A_1)\rho(A_1,A_2,t)] \nonumber\\
&&~~~~~~~~~~-\mu \rho(A_1,A_2, t) + 8 \mu \rho(A_1,A_2, t),\label{eq:dilution2}
\end{eqnarray}

The simulation considers a system where cells are initially in an
equilibrium distribution for a non-dividing population (solid curve in
Fig.~\ref{fig:divisionbias} for vertical axis, and day 0 density in
Fig.~\ref{fig:dye} for horizontal axis). Beginning on day 0 in this
simulation, the cells are stimulated to divide. This simulation has
correspondence with resting lymphocytes that are dyed with CFSE before
the system is infected on day zero, initiating rapid lymphocyte
division. Contours are drawn with logarithmic spacing. On day 7 shown
in Fig.~\ref{fig:dilutiondownreg}, $A_1$ is diluted as the population
divides. There has been no internal change in the chemical dynamics as
is typically considered in down-regulation of a gene-product. The use
of population-expression models can help to discriminate between a
down regulation where production rate $\alpha$ is decreased and where
simple dilution is occurring.

\subsubsection*{{\bf Cell division and spurious correlations:}}

Another effect of cell division is that two chemical quantities that
have independent dynamics can have correlations generated by cell
division. Cell division will cut both otherwise independent quantities
in half simultaneously. The population-expression equation is:
\begin{eqnarray}
\fl	\frac{\partial \rho(A_1,A_2, t)}{\partial t} &=&  -\frac{\partial}{\partial A_1} [(\alpha - \beta A_1)\rho(A_1,A_2,t)] + \frac{1}{2} \frac{\partial^2}{\partial
          A_1^2}[(\alpha + \beta A_1)\rho(A_1,A_2,t)] \nonumber\\
		&&-\frac{\partial}{\partial A_2} [(\delta - \epsilon A_2)\rho(A_1,A_2,t)] + \frac{1}{2} \frac{\partial^2}{\partial
          A_2^2}[(\delta + \epsilon A_2)\rho(A_1,A_2,t)] \nonumber\\
&&~~~~~~~~~~-2\mu \rho(A_1,A_2, t) + 8 \mu \rho(A_1,A_2, t),\label{eq:dilution3}
\end{eqnarray}
having a similar form to Eq.~(\ref{eq:dilution2}). However, $A_2$ also
obeys a simple gene-product rate law, and we include an extra factor of
2 in the second to last term for homeostatic division.

\begin{figure}[htbp]
	\centering
		\includegraphics[scale=1]{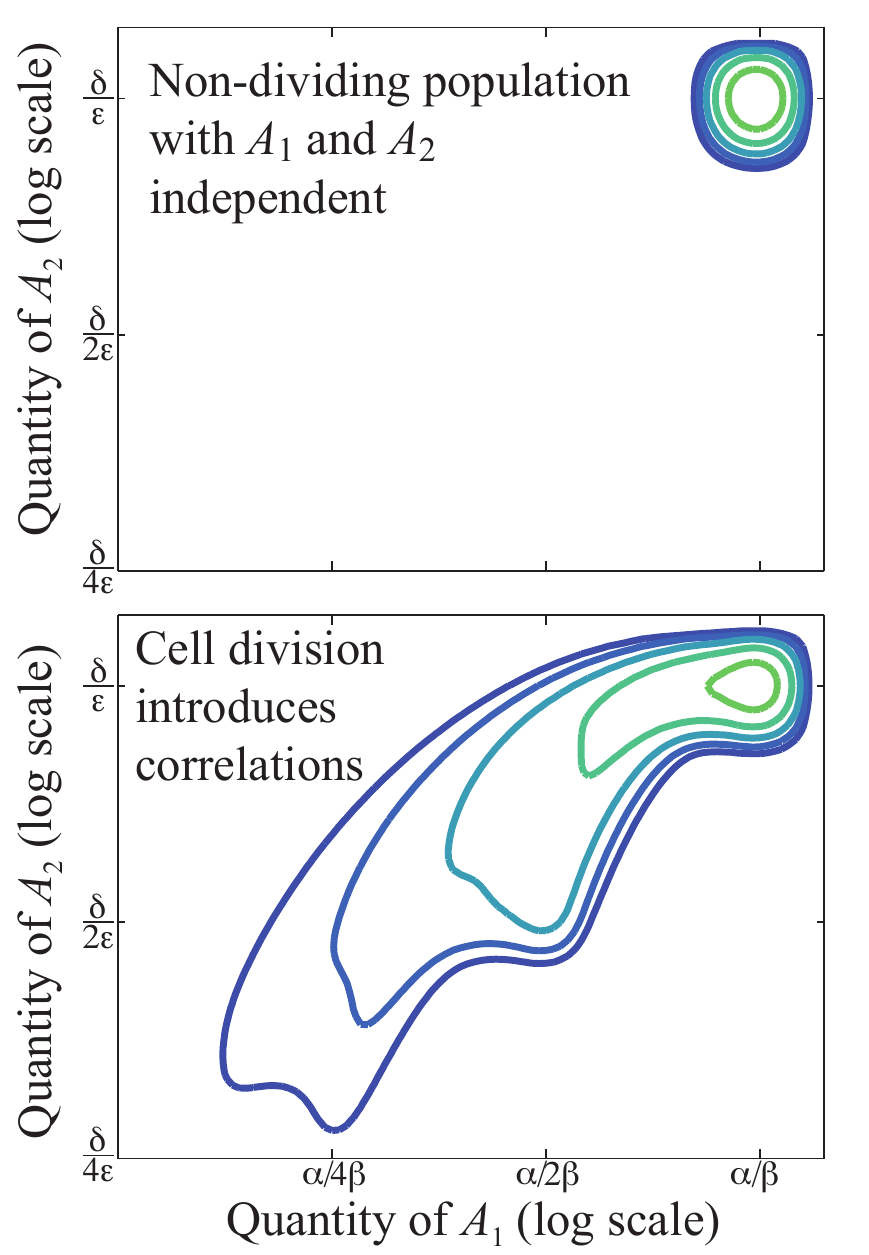}
                \caption{Cell division introduces correlations between
                  otherwise independent gene products. Here $A_1$ and
                  $A_2$ have simple rate laws ($dA_1/dt=\alpha-\beta
                  A_1$ and $dA_2/dt=\delta-\epsilon A_2$ with $A_2$
                  dynamics faster than $A_1$). In a non-dividing
                  population, we see that these products are not
                  correlated (top). In dividing cells, both $A_1$ and
                  $A_2$ are halved at the same time (cell division)
                  introducing correlations in expression level
                  (bottom). The effects of cell division should be
                  accounted for when analyzing expression data for
                  correlations to avoid spurious
                  conclusions. Numerical values used in this
                  simulation were $\alpha$ = 200 copies/day,
                  $\beta$=0.4 day$^{-1}$, $\delta$ = 800 copies/day,
                  $\epsilon$=1.6 day$^{-1}$, and $\mu$ = 0.07
                  day$^{-1}$ for the dividing population.}
	\label{fig:correlations}
\end{figure}

A simulation of the equilibrium distribution of
Eq.~(\ref{eq:dilution3}) is shown in Fig.~\ref{fig:correlations}. This
is a two-dimensional extension of Fig.~\ref{fig:divisionbias}. The
non-dividing population corresponding to the solid curve in
Fig.~\ref{fig:divisionbias} is depicted at top, and the correlated
spread resulting from cell division shown at bottom. The asymmetry in
the distribution is a result of the $A_2$ dynamics being more rapid
than the $A_1$ dynamics ($\delta>\beta$).

We note that correlated fluctuations in expression levels are
frequently used to infer the structure of genetic regulatory
\cite{Margolin:2006bc,stewart-ornstein_cellular_2012}, signaling
\cite{Sachs:2005kk}, and metabolic networks
\cite{Nemenman:2007kn}. Failing to account for the effects of cell
division in such an analysis can lead to the incorrect reconstruction
of the genetic network. Spurious correlations between gene-products
are strongest for pairs where both have slow degradation
rates. Correlations in gene-product expression result very naturally
from cell division. These correlations are typically grouped with
other forms of {\em extrinsic noise}
\cite{swain_intrinsic_2002}. Population-expression models allow us to
resolve the relative magnitude of different noise sources in extrinsic
noise, potentially improving genetic regulatory network reconstruction
methods.

\subsection{Failure of single cell systems biology: selection bias}
Consider now a two gene example with  influx and selection.  Here there
is an initial population of cells localized around
$(A_1^0,A_2^0)$. These cells have chemical dynamics such that at
$t=0$, $A_1$ begins to rapidly decrease and $A_2$ begins to gradually
increase. The population dynamics that underlies selection in this system 
arises from changes in the rate of division and death of cells in a manner 
dependent on the concentrations of $A_1$ and $A_2$ within the cell. We set the division rate proportional to $A_1$
and the death rate proportional to $A_2$. The system also has a
gradual influx of cells $\Gamma(A_1,A_2)$ entering the system around
$(A_1^0,A_2^0)$. Here we do not consider the effects of dilution with
cell division.

The system is described, using the vector notation, by:
\begin{eqnarray}
	\gamma_1&=&\alpha-\beta A_1,\\
	\gamma_2&=&\delta-\epsilon A_2,\\
	\frac{\partial \rho}{\partial t} &=& -\vec \nabla \cdot\left[ \vec \gamma\rho\right]+\vec \nabla \cdot(\mathbf{D}\nabla\rho) +dA_1\rho-dA_2\rho+\Gamma(A_1,A_2),
\end{eqnarray}
where we have used terms introduced in Section 2 and omitted the
dependence of terms on the quantities $A_1$ and $A_2$ for brevity.

\begin{figure}[htbp]
	\centering
		\includegraphics[scale=1]{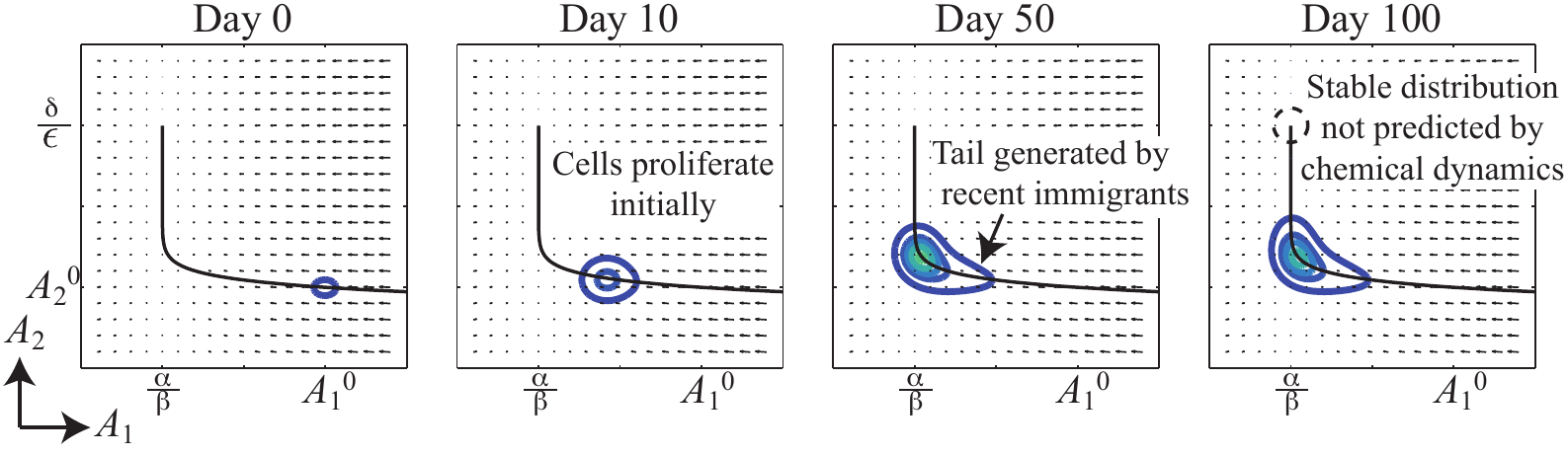}
                \caption{A two dimensional system with selection. Here
                  cells proliferate with rate proportional to $A_1$
                  and die with rate proportional to $A_2$. An initial
                  population localized around $(A_1^0,A_2^0)$ in the
                  lower right (Day 0) and proliferates as they begin
                  to differentiate (Day10). There is a constant but
                  gradual influx of new cells entering the system
                  around $(A_1^0,A_2^0)$ that can be seen biasing the
                  population by day 50, giving the population a tail
                  of higher $A_1$ expression. The $A_2$ dynamics are
                  slower and cells die before they ever reach the
                  steady state predicted by the chemical dynamics
                  (dashed black circle at Day 100). Instead a steady
                  state that is a product of the population dynamics
                  and the differentiation is reached which requires
                  constant influx to maintain. The black curve
                  illustrates the mean trajectory of the cells as
                  predicted by the chemical dynamics. In this
                  simulation $\alpha$ = 100 copies/day, $\beta$=1
                  day$^{-1}$, $\delta$ = 13.33 copies/day,
                  $\epsilon$=0.07 day$^{-1}$, $d$ = 0.04
                  copies$^{-1}$day$^{-1}$.}
	\label{fig:twoD}
\end{figure}

Fig.~\ref{fig:twoD} shows the evolution of the density in the
$A_1,A_2$ plane. We see that, by day 10, the initial population has
proliferated and progressed along the differentiation pathway. By day
50 we see there has been considerable proliferation and the cells that
are furthest along in the differentiation pathway have begun to
decay. We also see the effects of the gradual influx of new cells at
day 50 where the population now has a tail of recent immigrants that
have proliferated. By day 100 the initial population has decayed
completely and there is a stable distribution. This stable
distribution is maintained by the influx of new cells and not by a
lack of cell turnover. There is no change in ``phenotypic state'' for
the cells in this simulation, meaning cells maintain the same
production and degradation values for $A_1$ and $A_2$ throughout the
simulation.

The average differentiation path of a cell is given by the solid black
curve in Fig.~\ref{fig:twoD}. Though the stable distribution is
localized, it does not correspond to the phenotypic state described by
the $A_1$, $A_2$ dynamics. The population dynamics gives a strong bias
to the distribution that is not predicted by the chemical dynamics
alone. In statistical physics it is common to use the
``fluctuation-dissipation theorem'' to estimate model parameters from
the equilibrium distribution. Any such analysis of a dynamic
population must also take selective effects into account
\cite{lehner_fluctuation_2011}.

Rather than discuss the dynamics of $A_1$ and $A_2$ separately we can
discuss the differentiation of cells moving along the one dimensional
average path (black curve). To do this we introduce the variable $a$
where cells enter the system at $a=0$ and the differentiation pathway
takes them towards $a=1$; though, as seen in Fig.~\ref{fig:twoD} Day
100, they may never reach $a=1$. In this reduced model we also
neglect the stochastic effects and the only heterogeneity in the
system is due to the influx of new cells.

The one dimensional description is given by:
\begin{eqnarray}
	\gamma&=&\epsilon - \epsilon a,\\
	A_1(a) &=& \frac{\alpha}{\beta} - \left(\frac{\alpha}{\beta}-A_1^0\right)(1-a)^{\beta/\epsilon}\label{eq:A1ofa},\\
	A_2(a) &=& A_2^0 (1-a)+\frac{\delta}{\epsilon}a\label{eq:A2ofa},\\
	\frac{\partial \rho(a,t)}{\partial t} &=& -\frac{\partial}{\partial a}\left[ (\eta - \eta a)\rho(a,t)\right] +dA_1(a)\rho-dA_2(a)\rho+\Gamma(a=0).
\end{eqnarray}
In this one-dimensional model we still have a distribution of cells
since influx of $a=0$ cells gives diversity to the system. In the
absence of this influx we can describe the population with an ODE
model where the population has an internal variable (a
zero-dimensional approximation):
\begin{eqnarray}
	\frac{da}{dt}&=&\epsilon - \epsilon a,\\
	\frac{d X(t)}{d t} &=& dA_1(a)X(t)-dA_2(a)X(t).
\end{eqnarray}
where $A_1(a)$ and $A_2(a)$ are described by Eq.~(\ref{eq:A1ofa}) and
(\ref{eq:A2ofa}). This approach was recently used to describe the
exhaustion of CD8 T cells during a chronic infection where the
internal variable corresponded to the level of exhaustion in the
population and where thymic influx could be neglected
\cite{johnson_vaccination_2011}.

\section{Choosing the right variables}
\label{model-reduction}

Traditional flow cytometry interrogates large numbers of cells.  However the information from a single cell is limited by the spectral overlap of the
fluorescent dyes to measuring the concentration of about fifteen different molecules.
Soon, new techniques such as Cy-TOF
\cite{bendall_single-cell_2011} (which merges mass-spectrometry with
flow cytometry) will allow us to overcome this limitation and obtain simultaneous 
measurements of the concentration of hundreds of molecules at the single cell level. As the dimensionality
increases, the techniques of population-expression modeling become
computationally intractable. This necessitates dimensional reduction
and identification of ``key players'' among the measured molecular
expressions. At the same time, even as we measure more and more
quantities, some of the key players will still be omitted, forcing us
to look for such important missing links.

In the simplest case, the expression dynamics for all chemical species
in the system would be determined by a few key regulators, $B_\mu$,
i.e., $ {dA_i}/{dt}=\alpha_i(\vec{B}, A_i)-\beta_i(\vec{B}, A_i)+{\rm
  \eta}$, where $\alpha$ and $\beta$ are the production/degradation
functions, and $\eta$ is the noise term. $B_\mu$ can be an individual
chemical species, or more likely some function of many of the
individual expressions. The goal is to find the minimal set $\vec{B}$
from data, or to understand if the data does not provide sufficient
information to do so.

There is no single universal approach for dealing with
large-dimensional data that would solve both of these problems in the
immunological context. In fact the problems are not unique to
immunology, or even to biology.  Classic dimensionality reduction
techniques include Principal Components Analysis (PCA)
\cite{Jolliffe:2002}, Independent Components Analysis (ICA)
\cite{Lee:1998}, LASSO regression \cite{Tibshirani:1996}, and other
approaches that explicitly identify (locally) linear subspaces spanned
by data \cite{Blatt:1997,Tenenbaum:2000,Roweis:2000}. Many of these
would be problematic in immunology since they measure importance by
explained variance, which changes depending on the measurement units
used. For example in PCA, using the measured brightness or its
logarithm as the raw data may give very different results.  The
problem is solved elegantly with information-theoretic approaches,
which are manifestly reparameterization invariant \cite{Cover:2006}.

For this and related reasons, some of the most successful
dimensionality reduction approaches in quantitative cell biology (and
in computational neuroscience) have relied on information-theoretic
techniques. For example, finding pairs of genes with high mutual
information among their microarray mRNA expression profiles that
cannot be explained away by confounding effects of other regulatory
interactions uncovers ``minimal'' transcriptional regulatory networks
in cells as complex as human lymphocytes
\cite{Margolin:2006bc}. Higher order information-theoretic analyses
\cite{Margolin:2010eo} further disambiguate scenarios where simple
pairwise interactions do not explain the data and more
complex regulatory patterns are needed instead, (e.g.~two or more factors
regulating expression \cite{Wang:2009jy}). Similarly, searching for projections of the
combinatorially complex stimulus space that preserve the information
about the rate of spiking is one of the most powerful methods for
finding receptive fields of neurons from electrophysiology data
\cite{Sharpee:2004}. All of these approaches are special cases of the
rate-distortion framework \cite{Cover:2006}, where a
``small'' description of data is sought that nonetheless preserves the
information about the variable of relevance \cite{Cover:2006, Tishby:2000}.  The
balance between the amount of information kept and the model size is
controlled by the needs of the modeler and the data availability.

These methods should work for the context of immunology, but some
changes are needed. First, typical immunology flow cytometry
experiments make it hard to assay many different phenotypic or
temporal conditions, as is typically used for information-theoretic
analyses \cite{Margolin:2006bc,Sharpee:2004}. This limits the range of
variation of the data and can artificially reduce the values of the
measured information quantities. Luckily, as demonstrated in
\cite{Cheong:2011jp}, having many (tens of) thousands of single cell
measurements allows accurate information estimation in these
scenarios. However, it is {\em crucial} for the measurements to be of
a {\em very} high accuracy.

The second distinction of immunological data is that, in the
foreseeable future, the number of profiled quantities will be in the
hundreds, but not in the thousands, with cell surface molecules being
the easiest to profile. This leaves a possibility for missing key
regulators in the data sets. As was demonstrated recently
\cite{Cheong:2011jp}, information theoretic analysis can detect when
such important regulators are missing. This is done by observing that
a missing regulator induces complex statistical dependences among all
of its targets that cannot be explained by simple pairwise
correlations \cite{Margolin:2010eo}. While identification of such
missing regulators in a semi-automated fashion is possible
\cite{Wang:2009jy}, the smaller dimensionality of the immunological
data requires resetting the balance between the precision and the
recall.

The third, and the most fundamental, distinction of immunological data
is their population-expression nature. As illustrated in
Fig.~\ref{fig:correlations}, cell division and death introduces
spurious statistical relations among the measured
expressions. Distinguishing effects of regulation vs.\ population on
the interactions among the measured variables should be possible by
measuring the statistics of relations among physically non-interacting
variables in experimental data and in numerical simulations.

\begin{figure}[htbp]
	\centering
		\includegraphics[scale=1]{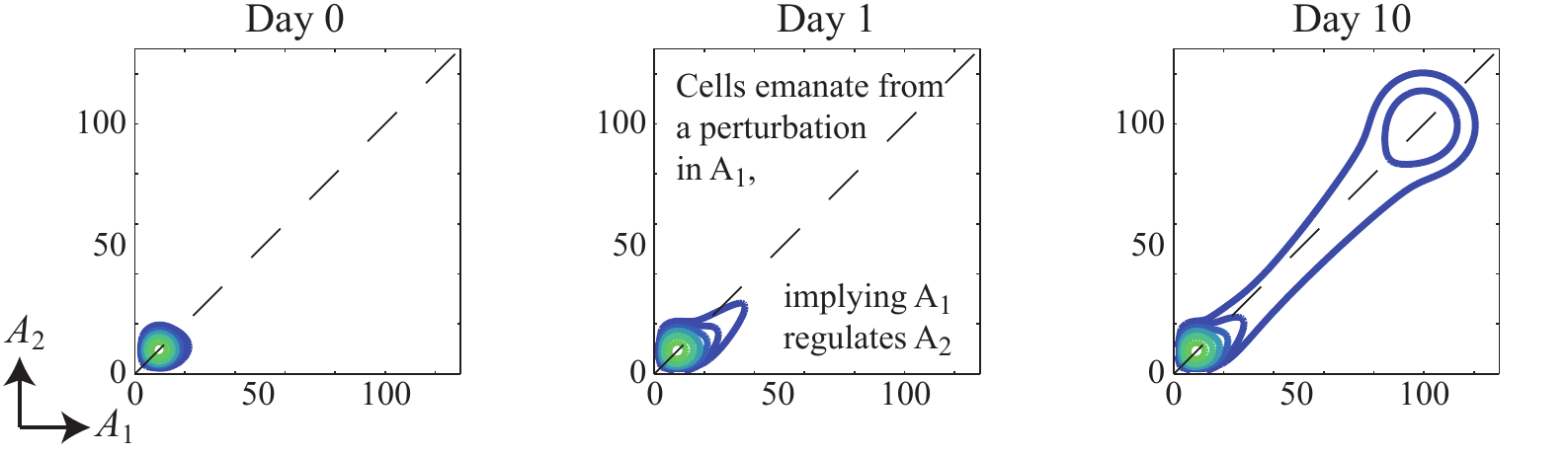}
                \caption{A two dimensional system of coupled
                  stochastic biochemical species, with deterministic
                  dynamics as in Eqs.~(\ref{eq:caus1}) and
                  (\ref{eq:caus2}), with $\alpha_0$ = 200 copies/day,
                  $\alpha_1$ = 2000 copies/day, $\beta$=22 day$^{-1}$,
                  $K$ = 30 copies, $n=6$. $A_1$ exhibits
                  bistability. Since it controls the expression of
                  $A_2$, the distribution of the latter is also
                  bimodal. Notice the asymmetry of the contour plots
                  of the joint probability distribution. By itself,
                  such asymmetry, as in the central panel, simply
                  signals unequal regulation of the two
                  species. However, time series measurements will
                  notice that the population average of $dA_2/dt$ is
                  correlated with the population average of $A_1$, but
                  not the other way around. Graphically, this
                  corresponds to the population escaping from the low
                  expression steady state along the $A_1$ direction
                  first, with $A_2$ following. This is a signal of the
                  potential causal regulation $A_1\to A_2$.}
	\label{fig:causality}
\end{figure} 

Since development and differentiation of immune cells is fast and can
be tracked in flow cytometry experiments on the scale of days, the
data offers an ability to establish causality of regulation
\cite{munsky_using_2012}.  This is in contrast to identification of
non-causal, symmetric relations among variables in most systems
biology or computational neuroscience data analysis approaches. We
illustrate this on the example of two coupled biochemical species
obeying the deterministic dynamics
\begin{eqnarray}
\frac{dA_1}{dt}=\alpha_0 +\frac{\alpha_1 A_1^n}{K^n+A_1^n}-\beta A_1,\label{eq:caus1}\\
\frac{dA_2}{dt}=\alpha_0 +\frac{\alpha_1 A_1^n}{K^n+A_1^n}-\beta A_2.\label{eq:caus2}
\end{eqnarray}
Here $A_1$ is self-regulating and can have two stable expression
levels. $A_2$ is regulated by $A_1$ and
will also be bimodal, but there's a clear difference between the two variables.
Solution of the corresponding Fokker-Planck system is shown in
Fig.~\ref{fig:causality}, illustrating that the dynamics of the transient shapes of
the joint probability distribution can signal the causality of
regulatory relations. 

This could be confirmed experimentally by sorting the cell population
at an early time (e.g.~day 1) into subpopulations based on expression
levels. The contrast between the dynamics of the $A_1$-high, $A_2$-low
subpopulation and the $A_1$-low, $A_2$-high subpopulation would reveal
which is the driver of the system. These sorted subpopulations would
be placed into animals where they are recognized by unrelated genetic
markers (e.g.~Thy1.1) and monitored to see which subpopulation reaches
$A_1$-high, $A_2$-high more rapidly.

\section{Conclusion}

Modeling in systems immunology is still in its infancy. Modeling
requires identifying the key players and parameters that describe the
behavior of interest. Population-expression models provide a tool for
interpreting the changing expression profiles of multi-cellular
populations that are differentiating while dividing and undergoing
selection. They achieve this by connecting the population scale with
intracellular systems biology.

The interpretation of immunological data has typically consisted of
enumerating cellular phenotypes and describing how the sizes of these
populations change over time. In contrast, the interpretation of data
with population expression models focuses on the chemical interaction
network common to all these phenotypes, and on the dependence of
expression levels on division and death rates. One could instead
continue adding additional phenotypic states to more accurately
describe the data, but this is reminiscent of the ``epicycles on
epicycles'' used to described the motion of the planets in the
Ptolemaic geocentric model of the universe. Looking at the problem
differently can yield both simplicity and insight.

A complete view of systems biology would capture population dynamics,
within-cell systems biology, and spatial effects. The spatial effects
like clustering can occur at different scales. At the within-cell scale for example, clustering of
molecules in the cell membrane plays an important role in the detection of antigen (infected cells) by T cells. At the population level, pathogens can be localized to the specific tissues and organs which they infect, while B and T cell responses occur in other sites such as the lymph nodes.  Some spatial effects can be easily incorporated into the population-expression framework. The
population-expression models are well suited to compartmentalization,
where one considers a population-expression equation for different
tissues and expression dependent trafficking rates between these
compartments. For finer scale spatial effects, the
population-expression approach breaks down, as the PDEs assume large
numbers of cells in the compartments. In these low density regimes one
must instead consider a model which treats cells discretely.  In
molecular systems biology, master equations and discrete stochastic
simulations using Gillespie and related algorithms are very commonly
used to describe the discreteness of stochastic changes in the
phenotype of individual cells \cite{Gillespie-07} alongside
continuous Fokker-Planck and Langevin equation approaches. For
methodological purposes, we built the current work around the
population-expression analogue of the Fokker-Planck equation. However,
it is clearly possible to develop the corresponding master equations
and stochastic simulation algorithms, where the number of cells in a
certain chemical state would be tracked. Nonlocal transitions due to
cell division and related phenomena are not conceptually difficult to
implement in such approaches, but the number of types of possible
transitions, and hence the time complexity of a simulation, might grow
excessively because of the nonlocality. We leave the development of
these simulation algorithms for future publications.

Advances in a field often require the integration of theoretical and
experimental approaches.  In the past the use of cellular dynamics
data, such as flow cytometric data, typically allowed us to enumerate
large numbers (millions) of cells but restricted us to making a
handful of measurements on each cell, limiting the phenotypic
resolution.  The extension of traditional flow cytometry to Cy-TOF
\cite{bendall_single-cell_2011} allows the measurement of hundreds of
biochemical species simultaneously at the single cell level.  This
allows, for the first time, tracking cellular systems biology dynamics
and the population dynamics simultaneously and with high accuracy.
The aim is to understand interactions among internal states of single
cells and the composition of cellular populations, and hence the
responses of the populations to infections. In this article we touched
upon key problems that need to be addressed for such analysis:
simultaneous representation of molecular system and population
dynamics, including proliferation and cell death, and
identification of key components of regulatory networks. We outlined a
few ways in which these problems can be tackled computationally, by
modifying current analysis approaches and by 
introducing population-expression modeling.

\section{Acknowledgements}
The authors would like to thank Rama S. Akondy and Raffi Ahmed for
Fig.~\ref{fig:cytometry} and helpful discussion. SPS was funded by NIH
grant U01 GM070749, RA was funded by NIH grant R01 AI049334, IN was
funded by HFSP grant RGY0084/2011,  NIH/NCI grant 7R01 CA132629-04, and James S.\
McDonnell Foundation Complex Systems Research Award 220020321. 

\pagebreak 

\section*{References}
\bibliographystyle{unsrt}
\bibliography{EnzymeReview}


\pagebreak
\renewcommand{\theequation}{S.\arabic{equation}}

\section*{Supplemental Derivations for Population-expression models of immune response}
There are a number of ways to derive Eq.~(2) of the main text\footnote{Throughout this Supplemental, references to equations in the main text are given without a prefix i.e.~ Eq.~(2), while references to equations found within the Supplemental are prefixed with S.}, or to derive the combination of the advection term in Eq.~(2) with the diffusion term of Eq.~(5). Several are presented in {\it The Dynamics of Physiologically Structured Populations} \cite{metz_dynamics_1986}. That volume is not in wide circulation, so for the interested reader we present two methods of deriving those equations. The first derivation is is more typical of fluid dynamics texts. In that field the result is typically called either an advection-diffusion equation or a convection-diffusion equation. The second derivation we provide utilizes a chemical master equation to derive a Fokker-Planck equation. This derivation is more typical to statistical physics and chemistry. The end results of both methods are equivalent. 

\section*{Divergence Theorem Derivation}
A typical within cell model specified by ordinary differential equations has the form:
\begin{eqnarray}
	\frac{dA_1}{dt}&=&\gamma_1(A_1,A_2,\ldots,A_n),\\
	\frac{dA_2}{dt}&=&\gamma_2(A_1,A_2,\ldots,A_n),\\
	~~\vdots && ~~~~~~~~~\vdots \nonumber\\
	\frac{dA_n}{dt}&=&\gamma_n(A_1,A_2,\ldots,A_n).
\end{eqnarray}

\begin{table}
\begin{center}
\begin{tabular}{p{13.0cm}}
\begin{shaded}
The example within-cell models used in the text were a dye, generating Eq.~(10):
\begin{equation}
	\frac{dA_1}{dt} = 0,
\end{equation}
simple protein production and decay, generating Eq.~(11):
\begin{equation}
	\frac{dA_1}{dt} = \alpha - \beta A_1,
\end{equation}
dye and protein, generating Eq.~(12):
\begin{eqnarray}
	\frac{dA_1}{dt}&=&\alpha - \beta A_1,\\
	\frac{dA_2}{dt}&=&0,
\end{eqnarray}
two non-interacting proteins, generating Eq.~(13) and (16):
\begin{eqnarray}
	\frac{dA_1}{dt}&=&\alpha - \beta A_1,\\
	\frac{dA_2}{dt}&=&\delta - \epsilon A_2,
\end{eqnarray}
and an autoregulator controlling itself and another protein, found in Eq.~(23) and (24):
\begin{eqnarray}
\frac{dA_1}{dt}&=&\alpha_0 +\frac{\alpha_1 A_1^n}{K^n+A_1^n}-\beta A_1,\\
\frac{dA_2}{dt}&=&\alpha_0 +\frac{\alpha_1 A_1^n}{K^n+A_1^n}-\beta A_2.
\end{eqnarray}
The first system of the paper had a discrete variable for transcription factor binding. In that case we had two equations for $A_1$ which generated Eq.~(6) and (7): 
\begin{equation}
	\frac{dA_1}{dt}=\alpha_{off}-\beta A_1,
\end{equation}
and 
\begin{equation}
	\frac{dA_1}{dt}=\alpha_{on}-\beta A_1.
\end{equation}
\end{shaded}
\end{tabular}
\end{center}
\caption{\label{tab:Struc1} Example within-cell models used in the text.
}
\end{table}

\begin{figure}[htbp]
	\centering
		\includegraphics[scale=1]{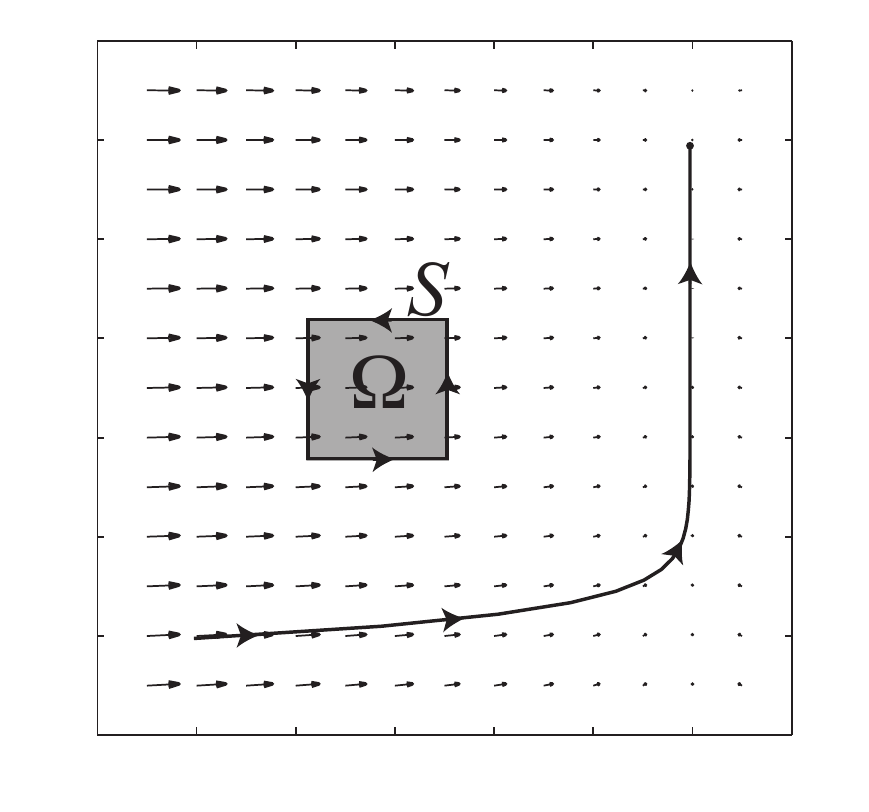}
	\caption{The arrows illustrate a possible velocity vector field in the chemical configuration space. The solid black curve illustrates the trajectory of a single cell going from low-low (vertical-horizontal) expression to low-high and finally to high-high. The shaded region $\Omega$ and its boundary $S$ are considered in the derivation of the dynamics of a density governed by this velocity field.}
	\label{fig:trajectory}
\end{figure}

These examples were all either one or two dimensional for simplicity, but as in Eqs.~(S.1-3), the general system is $n$ dimensional. As in the main text, the system of equations defines a velocity vector field $\vec\gamma(\vec A)$. We can view the chemical quantities $A_i$ as spatial variables rather than time-dependent quantities giving the trajectory of a cell moving through the space. A possible trajectory is shown in Fig.~\ref{fig:trajectory}. 

To derive the equation of motion for a density, denoted $\rho(\vec A,t)$, which is also governed by the vector field $\vec\gamma(\vec A)$, consider a volume $\Omega$ in the phase space (Fig.~\ref{fig:trajectory}). We ask how the total number of cells within this volume changes with time. We presently restrict the system to be conservative (no cell division or death) and deterministic (we relax these assumptions below). We can compare the time derivative of the total number of cells within the volume (spatial integral over $\Omega$) to the flux crossing the boundary ($S$):
\begin{eqnarray}
	\frac{d}{dt} \int_{\Omega} \rho(\vec{A},t) d\vec{A} + \oint_{S} \vec{n}\cdot \left[\vec{\gamma}(\vec{A})\rho(\vec{A},t)\right] dS &=& 0,\\
	\frac{d}{dt} \int_{\Omega} \rho(\vec{A},t) d\vec{A} + \int_{\Omega} \vec \nabla \cdot \left[\vec{\gamma}(\vec{A})\rho(\vec{A},t)\right] d\vec{A} &=& 0,\\
	\int_{\Omega} \left( \frac{\partial \rho}{\partial t} + \vec \nabla \cdot \left[\vec{\gamma}(\vec{A})\rho(\vec{A},t)\right] \right)d\vec{A} &=& 0,\\
	\frac{\partial \rho}{\partial t} = - \vec \nabla \cdot \left[\vec{\gamma}(\vec{A})\rho(\vec{A},t)\right],  &&\label{eq:Advec}
\end{eqnarray}
where $\vec{n}$ is the normal vector to the surface and the flux of cells at any point is the density times the velocity, $\rho(\vec{A},t)\vec{\gamma}$. The second line of this derivation utilizes the divergence theorem to replace an integral over the boundary with an integral over the enclosed region. The differential operator $\vec \nabla$ is defined as a vector of partial derivative operators ($\partial/\partial A_1,\partial/\partial A_2,\ldots$).In the last line we have made use of the fact that the integral is zero independent of the region defined. This yields an advection equation (sometimes called a convection equation) for describing the population expression of cells governed by $\vec{\gamma}$.

To this result we can append the non-conservative terms for cell division, death, migration and discrete variable changes. Appending these terms can be justified by noting that the creation rate within the volume element would be $(d\rho/dt)d\vec A$. Carrying that additional term through the derivation gives:
\begin{equation}
	\frac{\partial \rho}{\partial t} = - \vec \nabla \cdot \left[\vec{\gamma}(\vec{A})\rho(\vec{A},t)\right] + \frac{d\rho}{dt}.
\end{equation}

We can also extend our result using Fick's law which defines the flux resulting from stochastic fluctuations. Our system in general has position-dependent and direction-dependent fluctuations. This gives an inhomogeneous and anisotropic diffusion tensor $\mathbf{D}(\vec A)$. Fick's law, which we do not derive here, defines the flux as:
\begin{equation}
	\vec{J}_{\rm diff} = -\mathbf{D}(\vec A) \nabla \rho(\vec A,t).
\end{equation}
Adding this term in our above derivation and carrying through using the divergence theorem gives the standard form: 
\begin{eqnarray}
	\frac{d}{dt} \int_{\Omega} \rho(\vec{A},t) d\vec{A} + \oint_{S} \vec{n}\cdot \left[\vec{\gamma}(\vec{A})\rho(\vec{A},t)\right] dS && \nonumber\\
	 ~~~~~~~~~~~~~~~~~~~~~~~~~~~~~~~~~~~~~- \oint_{S} \vec{n}\cdot \left[ \mathbf{D}(\vec A) \nabla \rho(\vec A,t) \right] dS &=& 0,\\
	\frac{\partial \rho}{\partial t} = - \vec \nabla \cdot \left[\vec{\gamma}(\vec{A})\rho(\vec{A},t)\right] +\vec \nabla \cdot \left[\mathbf{D}(\vec{A})\nabla \rho(\vec{A},t)\right]. &&
\end{eqnarray}

\section*{Chemical Master-Equation Derivation}
A second method of deriving the advection-diffusion equation starts with a chemical master equation. A comprehensive introduction to master equation techniques can be found in {\it Stochastic processes in physics and chemistry} \cite{kampen_stochastic_2007}. The formalism for the general chemical master equation is cumbersome, but available, and the reference material provides methods for deriving the advection-diffusion equation for general chemical systems. Rather than provide the general derivation we present the derivation of the advection-diffusion equation for the two independent products in Eq.~(13) and (16). 

For that system we have four reactions: production of $A_1$ with rate $\alpha$, degradation of $A_1$ with rate $\beta A_1$, production of $A_2$ with rate $\delta$, and reduction of $A_2$ with rate $\epsilon$. The master equation (in actuality an infinite number of coupled ordinary differential equations) describes the probability $P(A_1,A_2,t)$, of having $A_1$ and $A_2$ molecules of each species at time $t$. In a chemical master equation the quantities $A_i$ are treated as discrete variables. With the described reactions and rates we have the master equation:
\begin{eqnarray}
	\frac{d P(A_1,A_2,t)}{dt}&=&~~~~\alpha P(A_1-1,A_2,t)- \alpha P(A_1,A_2,t)\nonumber\\
	&&+\beta (A_1+1) P(A_1+1,A_2,t)- \beta A_1 P(A_1,A_2,t)\nonumber\\
	&&+\delta P(A_1,A_2-1,t)- \delta P(A_1,A_2-1,t)\nonumber\\
	&&+\epsilon (A_2+1) P(A_1,A_2+1,t)- \epsilon A_2 P(A_1,A_2,t).
\end{eqnarray}
It is helpful to define the step-operator $\mathbb{E}_i$ which operates on $A_i$ raising it to $A_i+1$ and likewise the inverse operator $\mathbb{E}_i^{-1}$ which lowers it. This gives us:
\begin{eqnarray}
	\frac{d P(A_1,A_2,t)}{dt}&=&~~~~\alpha (\mathbb{E}_1^{-1}-1)P(A_1,A_2,t)
	+\beta (\mathbb{E}_1-1) A_1 P(A_1,A_2,t)\nonumber\\
	&&+\delta (\mathbb{E}_2^{-1}-1) P(A_1,A_2,t)\nonumber\\
	&&+\epsilon (\mathbb{E}_2-1) A_2 P(A_1,A_2,t).
\end{eqnarray}

If we extend the chemical quantities to be a continuous variables we can obtain an approximate form for the step operator by noting that:
\begin{equation}
	f(A+1) = f(A) + f'(A) + \frac{1}{2!}f''(A) + \ldots,
\end{equation}
giving:
\begin{equation}
	\mathbb{E} = 1 + \frac{\partial}{\partial A} + \frac{1}{2!}\frac{\partial^2}{\partial A^2} + \ldots.
\end{equation}
Using the approximate forms we obtain:
\begin{eqnarray}
	\frac{d P(A_1,A_2,t)}{dt}&\approx& \left(-\frac{\partial}{\partial A_1} + \frac{1}{2}\frac{\partial^2}{\partial A_1^2}\right) \alpha P(A_1,A_2,t) \nonumber\\
	&&+\left(\frac{\partial}{\partial A_1} + \frac{1}{2}\frac{\partial^2}{\partial A_1^2}\right) \beta A_1 P(A_1,A_2,t)\nonumber\\
	&&+\left(-\frac{\partial}{\partial A_2} + \frac{1}{2}\frac{\partial^2}{\partial A_2^2}\right) \delta P(A_1,A_2,t)\nonumber\\ 
	&&+ \left(\frac{\partial}{\partial A_2} + \frac{1}{2}\frac{\partial^2}{\partial A_2^2}\right) \epsilon A_2 P(A_1,A_2,t),
\end{eqnarray}
which can be rewritten:
\begin{eqnarray}
	\frac{d P(A_1,A_2,t)}{dt}&\approx& -\frac{\partial}{\partial A_1} \left[(\alpha-\beta A_1) P(A_1,A_2,t)\right] \nonumber\\ 
	&&-\frac{\partial}{\partial A_2} \left[(\delta-\epsilon A_2) P(A_1,A_2,t)\right] \nonumber\\
	&& + \frac{1}{2}\frac{\partial^2}{\partial A_1^2}\left[ (\alpha + \beta A_1) P(A_1,A_2,t) \right] \nonumber\\
	&&+ \frac{1}{2}\frac{\partial^2}{\partial A_2^2}\left[ (\delta + \epsilon A_2) P(A_1,A_2,t) \right].
\end{eqnarray}
In vector notation with $\gamma_1=\alpha-\beta A_1$, $\gamma_2=\delta-\epsilon A_2$, and the diffusion tensor $\mathbf{D}(\vec A)$ given by:
\begin{equation}
	\mathbf{D}(\vec A) = \left[
	\begin{array}{cc}
		\frac{1}{2}(\alpha + \beta A_1) & 0\\
		0 & \frac{1}{2}(\delta + \epsilon A_2)
		\end{array},
	\right]	
\end{equation}
we have the advection-diffusion equation given by:
\begin{equation}
		\frac{\partial \rho}{\partial t} = - \vec \nabla \cdot \left[\vec{\gamma}(\vec{A})\rho(\vec{A},t)\right] + \vec \nabla \cdot \left[\mathbf{D}(\vec{A})\nabla \rho(\vec{A},t)\right], 
\end{equation}
where we have also replaced single-cell probability $P$, with the density $\rho$. Our diffusion is inhomogeneous, having a dependence on $\vec A$, and anisotropic, having different noise amplitudes in different directions, giving a diffusion tensor rather than a diffusion constant. In general the diffusion tensor will not be diagonal.

\end{document}